\documentclass[letterpaper,titlepage,11pt]{article}
\usepackage{hyperref}

\usepackage{amssymb,amsmath,amsfonts}
\usepackage{epsfig}

\def\clock{{\count0=\time
           \divide\count0 60
           \ifnum\count0<10 0\fi\the\count0
           \multiply\count0 -60 \advance\count0 \time
           :\ifnum\count0<10 0\fi \the\count0
         }}
\newcommand{\timestamp}{{\small\vbox{\hbox{\tt\jobname.tex}
\hbox{\the\day/\the\month/\the\year, \clock}}}}

\newcommand{\be}{\begin{equation}} \newcommand{\ee}{\end{equation}}
\newcommand{\bea}{\begin{eqnarray}} \newcommand{\eea}{\end{eqnarray}}

\newcommand{\CO}{\mathcal{O}}

\newcommand{\CT}{\mathcal{T}}

\newcommand{\CM}{\mathcal{M}}

\newcommand{\id}{\hbox{1\kern-.27em l}}
\newcommand{\sid}{\hbox{\scriptsize1\kern-.27em l}}

\newcommand{\we}{\kern-.1em\wedge\kern-.1em}
\newcommand{\scal}{\kern-.13em\cdot\kern-.13em}

\newcommand{\II}{I\kern-.09em I}

\newcommand{\R}{\mathbb{R}}

\newcommand{\nn}{\nonumber}

\newcommand{\spa}{\ , \ \ }

\newcommand{\rht}{\tilde{\rho}}
\newcommand{\tht}{\tilde{\theta}}

\numberwithin{equation}{section}

\begin{document}

\begin{titlepage}

\rightline{\vbox{\small\hbox{\tt hep-th/0310259} }}
\vskip 3cm

\centerline{\Large \bf Small Black Holes on Cylinders}

\vskip 1.6cm
\centerline{\bf Troels Harmark}
\vskip 0.5cm
\centerline{\sl The Niels Bohr Institute}
\centerline{\sl Blegdamsvej 17, 2100 Copenhagen \O, Denmark}

\vskip 0.5cm

\centerline{\small\tt harmark@nbi.dk}

\vskip 1.6cm

\centerline{\bf Abstract} \vskip 0.2cm 
\noindent
We find the metric of small black holes on cylinders,
i.e. neutral and static black holes with a small mass 
in $d$-dimensional Minkowski-space times a circle.
The metric is found using an ansatz for black
holes on cylinders proposed in hep-th/0204047.
We use the new metric to compute corrections to the thermodynamics
which is seen to deviate from that of 
the $(d+1)$-dimensional Schwarzschild black hole.
Moreover, we compute the leading correction 
to the relative binding energy which is found to be non-zero. 
We discuss the consequences of these results 
for the general understanding of black holes and 
we connect the results to the phase structure
of black holes and strings on cylinders.


\end{titlepage}


\pagestyle{plain}
\setcounter{page}{1}
\tableofcontents

\section{Introduction}

Neutral and static black holes on cylinders%
\footnote{With a black hole on a cylinder $\R^{d-1} \times S^1$
we mean a black hole in a $d+1$ dimensional space-time 
that asymptotes to $\CM^d \times S^1$ far away from the black hole,
where $\CM^d$ is the $d$-dimensional Minkowski space-time.}
$\R^{d-1} \times S^1$ 
have a more interesting dynamics and richer phase structure
than black holes on flat space $\R^d$. 
Neutral and static black holes in flat space 
are for a given mass $M$ uniquely described 
by the Schwarzschild solution with mass $M$.
With black holes on cylinders we can form a 
dimensionless quantity since the radius of the cylinder
gives us an extra macroscopic scale in the system.
This means that the behavior of the black holes
can depend highly on the value of such a dimensionless
quantity. 
This ties together with the fact that the cylinder
$\R^{d-1} \times S^1$ has a non-trivial topology
since it has a non-contractible cycle. 
This makes it possible for the black hole to grow
so big that its event horizon can ``meet itself'' across
the cylinder. It moreover makes it possible to have
other types of black objects like for example black
strings which have an event horizon that wrap across the
cylinder.

That the phase structure of black
holes on cylinders are richer than for example the Schwarzschild 
black holes can also be tributed to the fact that
the cylinder space-time $\CM^d \times S^1$
is not maximally symmetric unlike 
the Minkowski, de-Sitter and Anti-de-Sitter
space-times. Until now, black hole solutions have only
been found for maximally symmetric space-times or for
other highly symmetric space-times.
In particular solutions describing 
black holes on $\R^2 \times S^1$ have
been found \cite{Myers:1987rx,Bogojevic:1991hv,Korotkin:1994dw,Frolov:2003kd}
using the Israel-Khan solution \cite{Israel:1964}. 
However, the $\R^2 \times S^1$ cylinder is very different
from the $\R^{d-1} \times S^1$ cylinders for $d \geq 4$ since
for the $\CM^{3} \times S^1$ space-time there are enough killing
vectors to find the solution using the construction of 
Weyl \cite{Weyl:1917}. For $\CM^d \times S^1$ with $d \geq 4$
there are instead too few killing vectors to find a Weyl solution
\cite{Emparan:2001wk} which also is reflected in the fact that 
the metric for black holes on such cylinders does not belong to
an algebraically special class \cite{DeSmet:2002fv}.

The rich phase structure for black objects on cylinders
have been explored from many points of view. 
Gregory and Laflamme \cite{Gregory:1993vy,Gregory:1994bj} 
discovered that uniform black strings
on cylinders, i.e. strings that are wrapped symmetrically
around the cylinder, are unstable to linear perturbations 
when the mass of the string is below a certain critical mass.
This was interpreted to mean that a light uniform black string
decays to a black hole on a cylinder since that has higher 
entropy. However, Horowitz and Maeda \cite{Horowitz:2001cz} argued 
that this transition should have an intermediate step in the
form of a light non-uniform black string. 
Such a non-uniform string branch have not been found, but
a new branch of non-uniform strings have 
been found by \cite{Gregory:1988nb,Gubser:2001ac,Wiseman:2002zc}. 
This new branch of non-uniform strings 
seemingly does not exists for the mass range when
the uniform string is classically unstable.

Several proposals for the phase structure of black objects on 
cylinders 
have been put forward \cite{Horowitz:2001cz,Gubser:2001ac,Harmark:2002tr,Horowitz:2002dc,Kol:2002xz,Wiseman:2002ti,Harmark:2003fz,Kol:2003ja,Harmark:2003dg,Harmark:2003eg,Kol:2003if,Sorkin:2003ka,Kudoh:2003ki}.%
\footnote{Other recent and related work includes 
\cite{Casadio:2000py,Casadio:2001dc,Horowitz:2002ym,Kol:2002dr,Sorkin:2002nu,Choptuik:2003qd,Emparan:2003sy}.}
In \cite{Harmark:2003dg,Harmark:2003eg} 
a new phase diagram, the $(M,n)$ phase diagram,
was proposed as a tool to understand the phase structure
of black objects on cylinders. A similar
proposal for a phase diagram was made in \cite{Kol:2003if}.
One can thus formulate the main goal of this field
of research as follows: To draw the complete $(M,n)$
phase diagram depicting all the possible phases of black
objects on cylinders.

Finding the solution for black holes on cylinders
is therefore part of the larger question of understanding
the phase structure of black objects on cylinders.
Numerical studies 
of black holes on the cylinder $\R^4 \times S^1$ was recently
done in \cite{Sorkin:2003ka,Kudoh:2003ki},
but it is nevertheless still desirable to get a better analytic
understanding of black holes on cylinders in order
to provide definite answers to the questions regarding
the phase structure of black holes on cylinders.

Progress towards finding a solution for black holes on cylinders
was made in \cite{Harmark:2002tr} where an ansatz was proposed for metrics
describing black holes on cylinders $\R^{d-1} \times S^1$
with $d \geq 4$. In \cite{Harmark:2003eg} it was proven
that any neutral and static 
black hole on a cylinder can be put in this ansatz.
The proof was a generalizing of a proof of Wiseman \cite{Wiseman:2002ti}.
However, even though the ansatz of \cite{Harmark:2002tr} is highly
constrained the equations of motion are still very hard to solve.

In this paper we consider therefore the more tractable problem
of finding a metric describing {\sl small} black holes on cylinders,
i.e. black holes on cylinders with a small mass.
We use the ansatz of \cite{Harmark:2002tr} and find 
a metric describing the complete small black hole space-time, from
the horizon to the asymptotic region far away from the black
hole. 

We use our new solution for small black holes on cylinders to 
find the corrected thermodynamics. The thermodynamics becomes that of 
a Schwarzschild black hole in $d+1$ dimensions when the
mass goes to zero. But it deviates from the 
Schwarzschild black hole thermodynamics once the mass is non-zero.

We find furthermore the relative binding energy and
draw the $(M,n)$ phase diagram with the black hole branch
in the case $d=5$. 

The structure of the paper is as follows:
In Section \ref{secprel} we introduce the basic tools necessary
for constructing the small black hole solution.
In Section \ref{asympt} we first review the 
measurements of asymptotic quantities of \cite{Harmark:2003dg}.
We then review in Section \ref{secans} the ansatz of \cite{Harmark:2002tr} 
for black holes on cylinders.
In Section
\ref{fourier} the Fourier modes of the black hole branch are found.
Using this, we obtain 
the flat-space limit of the black holes on
cylinders in the specific ansatz.

In Section \ref{secnewcoords} we modify the ansatz by changing the
coordinates. This proves useful for constructing the small black
hole solution.  
After defining the new coordinates we subsequently
consider the flat space limit of the ansatz with
the new coordinates.
In Appendix \ref{apptherm} the thermodynamics of both
coordinate systems are considered.

In Section \ref{secNewt} we take the first step towards constructing
the small black hole solution by finding the first correction
to flat space far away from the black hole. 
In Section \ref{secn0sol} we find the correction by first
considering an arbitrary Newtonian gravity potential and then
using the result for the specific black hole case.
In Section \ref{secinrt} we then transform the result to the 
ansatz in the new coordinate system.

In Section \ref{secfull} we use the results of Section \ref{secNewt} to
find the complete metric for small black holes on cylinders.
We also use the results of Appendix \ref{appsph} where
general spherical metrics are considered.

In Section \ref{cortherm} we use the metric of Section \ref{secfull}
to find the corrected thermodynamics of black holes on cylinders.
This is used in Section \ref{secdiag} to draw the $(M,n)$ phase
diagram for black holes and strings on cylinders in the $d=5$ case.

Finally, we conclude the paper in Section \ref{secconcl}.

\subsubsection*{Note added}

Results on small black holes on cylinders that overlap with
the results of this paper were announced in \cite{Kol:2003if,Sorkin:2003ka}
to appear in the near future in a paper of D. Gorbonos and B. Kol.

\section{Preliminaries}
\label{secprel}

In this section we lay the groundwork necessary to construct
the corrected black hole on cylinder solution.
In Section \ref{asympt} we review how the asymptotically measurable
quantities are defined.
In Section \ref{secans} we present the general ansatz for the metric
of black holes on cylinders.
In Section \ref{fourier} we give argue what the Fourier modes
of black holes on cylinders should be and
use this to describe the flat-space
limit of the ansatz for the metric in detail.

\subsection{Asymptotically measurable quantities}
\label{asympt}

In \cite{Harmark:2003dg,Harmark:2003eg}
a program was set forth to cathegorize all static
vacuum solutions of higher-dimensional General Relativity 
(i.e. pure gravity solutions)
that asymptotes to $\CM^d \times S^1$, i.e.
all black objects on the cylinder $\R^{d-1} \times S^1$, 
according to their asymptotic
behavior.
In this section we review the ideas and results of 
\cite{Harmark:2003dg,Harmark:2003eg}
that are relevant to this paper.

In the following we define the physical parameters
that one can measure for any solution asymptoting to
$\CM^d \times S^1$. 
We parameterize here the 
metric for the flat space-time $\CM^d \times S^1$ as
\begin{equation}
\label{cylcoord}
ds^2 = -dt^2 + dr^2 + r^2 d\Omega_{d-2}^2 + dz^2 \ ,
\end{equation}
with $t$ being
the time, $r$ the radial coordinate in the $\R^{d-1}$ part and
$z$ the coordinate for $S^1$ with period $L = 2\pi R_T$.

In the rest of the paper we put $R_T=1$ (so that $L=2\pi$) 
to simplify our expressions.
Thus, $r$ and $z$ are dimensionless in the following, 
i.e. $r_{\rm new} = r_{\rm old}/R_T$ and $z_{\rm new} = z_{\rm old}/R_T$.
Moreover, $z$ has period $2\pi$ below.

To define our asymptotically measurable parameters we consider
Newtonian matter with energy momentum tensor
\begin{equation}
T_{00} = \rho \spa T_{zz} = - b \ .
\end{equation}
We define the mass $M$ and the relative binding energy $n$ by
\begin{equation}
M = \int d^d x \, \varrho (x) \spa
n = \frac{1}{M} \int d^d x \, b(x) \ .
\end{equation}
Note that we can use $n$ to define the
tension $\CT = nM/L$, which is the tension a string would have if
one had a string with same $M$ and $n$ as the black hole.
This is used as an alternative parameter to $n$ in \cite{Kol:2003if}.
See also \cite{Traschen:2001pb,Townsend:2001rg} 
for measurements of the tension $\CT$.

We define furthermore the two gravitational potentials
\begin{equation}
\label{pots}
\nabla^2 \Phi = 8\pi G_{\rm N} \frac{d-2}{d-1} \varrho
\spa
\nabla^2 B = - \frac{8\pi G_{\rm N}}{d-1} b \ ,
\end{equation}
where $G_{\rm N}$ is the $d+1$ dimensional Newtons constant.
Due to the conservation of the energy-momentum tensor
we require that $\partial_z b = 0$. This means that
$b = b(r)$, i.e. $b$ only depends on $r$.
Away from the mass-distribution we have then%
\footnote{Here 
$\Omega_k = 2 \pi^{(k+1)/2} / \Gamma ( \frac{k+1}{2} )$ 
is the volume of a unit $k$-sphere.}
\begin{equation}
\label{phirz}
\Phi(r,z) = - \frac{d-2}{(d-1)(d-3)} 4 G_{\rm N} 
\sum_{k=0}^\infty \frac{h( kr )}{r^{d-3}}  \cos( kz ) \varrho_k \ ,
\end{equation}
\begin{equation}
\label{brz}
B(r,z) = \frac{1}{(d-1)(d-3)} \frac{4 G_{\rm N}}{\Omega_{d-2}}
\frac{nM}{r^{d-3}}  \ ,
\end{equation}
with
\begin{equation}
\label{defh}
h(x) = 2^{-\frac{d-5}{2}} \frac{1}{\Gamma \left( \frac{d-3}{2} \right)}
x^{\frac{d-3}{2}} K_{\frac{d-3}{2}} (x) \ ,
\end{equation}
where $K_s(x)$ is one of the modified Bessel functions of the second kind
(in standard notation).
The coefficients 
$\varrho_k$, $k \geq 0$, are the Fourier modes of the mass-distribution.
Clearly, $\varrho_0 = M / \Omega_{d-2}$.

From the above we see that for an arbitrary static
mass-distribution of Newtonian matter
on $\CM^d \times S^1$ which is spherically symmetric
on $\R^{d-1}$ the measurable parameters are the mass
$M$, the relative binding energy $n$, and the Fourier modes
$\varrho_k$, $k \geq 1$.
We now turn to how to measure these parameters.

Independently of the gauge, we have that the $g_{tt}$ component
of the metric to first order in $G_{\rm N}$ is \cite{Harmark:2003dg}
\begin{equation}
\label{gtt}
g_{tt} = - ( 1 + 2\Phi + 2B) \ .
\end{equation}
If we work in a coordinate system where the leading correction to 
$g_{zz}$ for $r \rightarrow \infty$ is independent of $z$, we moreover have
that  \cite{Harmark:2003dg}
\begin{equation}
\label{gzz}
g_{zz} = 1 + \frac{1}{(d-1)(d-3)}\frac{4 G_{\rm N}}{\Omega_{d-2}}
\frac{(1-(d-2)n)M}{r^{d-3}} + \CO ( r^{-2(d-3)} ) \ ,
\end{equation}
is the leading correction to $g_{zz}$ for $r \rightarrow \infty$.
Therefore, using \eqref{gtt} and \eqref{gzz} we see that for
any given static metric $M$, $n$ and $\varrho_k$, $k\geq 1$, 
can be measured.

In particular, 
we define the mass $M$, the relative binding
energy $n$ and the Fourier modes $\varrho_k$, $k\geq 1$, 
for any static pure gravity solution asymptoting to $\CM^d \times S^1$
as what we measure by applying 
\eqref{gtt}-\eqref{gzz} with \eqref{phirz}-\eqref{defh}.%
\footnote{Notice that the measurements of the physical
quantities associated with the sources of the gravitational field
for solutions with event horizons are defined
in analogy with the results for non-gravitational Newtonian
matter. The reason behind this is the principle that any source of
gravitation affecting the asymptotic region the same way should
also have the same values for the physical parameters 
associated with
the sources of the gravitational field.}

We apply these results on the black hole on cylinder solutions
below.

\subsection{Ansatz for black hole solution}
\label{secans}

In order to find a metric for black holes on cylinders
$\R^{d-1} \times S^1$ it is important first to find an
ansatz for the metric that only has a limited number
of free functions. 
Progress in this direction were done in \cite{Harmark:2003eg} where
it was shown that the metric 
for any neutral and static black hole
on a cylinder $\R^{d-1} \times S^1$ which is spherically
symmetric on $\R^{d-1}$ can be written
in the form
\begin{equation}
\label{ansatz}
ds^2 = - f dt^2 + \frac{A}{f} dR^2 
+ \frac{A}{K^{d-2}} dv^2 + KR^2 d\Omega_{d-2}^2 
\spa
f = 1 - \frac{R_0^{d-3}}{R^{d-3}} \ ,
\end{equation}
where $A(R,v)$ and $K(R,v)$ are two functions specifying
the solution.
The ansatz \eqref{ansatz} was proposed in \cite{Harmark:2002tr}
for black holes on cylinders and was proven to be correct
in \cite{Harmark:2003eg} generalizing a proof
of Wiseman in \cite{Wiseman:2002ti}.

The properties of the ansatz \eqref{ansatz} was extensively 
considered in \cite{Harmark:2002tr}. 
It was found that $A(R,v)$ can be written explicitly
in terms of $K(R,v)$ thus reducing the number of unknown
functions to one. 
The functions $A(R,v)$ and $K(R,v)$ are periodic in $v$
with period $2\pi$.
Note that $R = R_0$ defines the location of the event
horizon for the black hole. 

The asymptotic region, i.e. the region far away from the black hole,
is located at $R \rightarrow \infty$. We impose the conditions
that $r/R \rightarrow 1$ and $z/v \rightarrow 1$ for $R \rightarrow
\infty$. This also means that $A,K \rightarrow 1$ for 
$R \rightarrow \infty$. 

We review the thermodynamics of the ansatz \eqref{ansatz}
in Appendix \ref{apptherm}.

As explained in \cite{Harmark:2002tr} and
in the introduction, finding a solution to
the equations for $A(R,v)$ and $K(R,v)$ is very hard. The equations
seems highly non-linear and so far no simplifications have been
found.
However, if we consider small black holes on cylinders, i.e.
small masses $M$, the equations simplify, as we shall see in
the following. We therefore
focus in the following on solving the equations for 
$A(R,v)$ and $K(R,v)$ to leading order in $R_0$ (the
$R_0 \rightarrow 0$ limit is equivalent to
the $M \rightarrow 0$ limit since
$M$ is proportional to $R_0^{d-3}$).

\subsection{Finding the Fourier modes and the flat-space limit}
\label{fourier}

We now consider the limit $M \rightarrow 0$ for a black
hole on a cylinder. 
Physically, it is clear that for very small masses the black
hole should behave as a point particle as seen from an observer
standing away from the black hole in the weakly curved region of
space-time. 
Thus, as $M \rightarrow 0$ the Newtonian potential
$\Phi(r,z)$ should become that of point-masses on a cylinder.
By the same token the relative binding energy $n$ should go
to zero, since the interaction of the black hole with itself
across the cylinder
becomes smaller and smaller as the black hole becomes smaller
(see also \cite{Harmark:2002tr} for a quantitative discussion of this).
We thus get that for $M \rightarrow 0$ the Newtonian potential is
\begin{equation}
\label{smallMphi}
\Phi(r,z) = - \frac{8\pi G_{\rm N} M}{(d-1)\Omega_{d-1}} F(r,z) \ ,
\end{equation}
with $F(r,z)$ given as
\begin{equation}
\label{defF}
F(r,z) = \sum_{k=-\infty}^{\infty} 
\frac{1}{(r^2 + (z-2\pi k)^2)^{\frac{d-2}{2}}} \ .
\end{equation}
Moreover, $B(r,z)/(G_{\rm N} M) \rightarrow 0$ for $M \rightarrow 0$
since $n$ should go to zero.
Note that we assume the black hole singularity to be located
at $(r,z)=(0,0)$.

The potential \eqref{smallMphi} is easily found using Newtons
law of gravity for points particles (use for example
\eqref{pots}).
Thus, the only thing we have used here is that for $M \rightarrow 0$
Newtons law of gravity
governs almost all of the space-time, except the vanishingly small
part close to the black hole, i.e. around $(r,z)=(0,0)$.

We can expand $F(r,z)$ in Fourier modes as
\begin{equation}
\label{expF}
F(r,z) = \frac{k_d}{r^{d-3}} \left( 1 + 2 \sum_{k=1}^{\infty}
h(kr) \cos(kz) \right) \ ,
\end{equation}
where $h(x)$ is given by \eqref{defh} and where we defined
\begin{equation}
k_d = \frac{1}{2\pi} \frac{d-2}{d-3} \frac{\Omega_{d-1}}{\Omega_{d-2}} \ .
\end{equation}
Using then \eqref{expF} we see that we can find the Fourier modes
$\varrho_k$ of $\Phi(r,z)$ in the $M \rightarrow 0$ limit.

We now consider the consequence of this observation for
black hole solutions in the ansatz \eqref{ansatz}.
Taking the $M \rightarrow 0$ limit is clearly the same as
taking the $R_0 \rightarrow 0$ limit. Define 
\begin{equation}
\label{defAK}
A_0 (R,v) = \lim_{R_0 \rightarrow 0} A(R,v)
\spa
K_0 (R,v) = \lim_{R_0 \rightarrow 0} K(R,v) \ .
\end{equation}
We then see from the ansatz \eqref{ansatz} and from \eqref{gtt} that
as consequence of \eqref{smallMphi} we get
\begin{equation}
\lim_{R_0 \rightarrow 0} \frac{R_0^{d-3}}{r^{d-3}} K_0(r,z)^{\frac{d-3}{2}}
= \lim_{R_0 \rightarrow 0} \frac{16 \pi G_{\rm N} M}{(d-1)\Omega_{d-1}} F(r,z)
\ .
\end{equation}
Since from \eqref{nt1} we have $M = 
\frac{\Omega_{d-2}}{8 G_{\rm N}} \frac{(d-1)(d-3)}{d-2} R_0^{d-3}$
in the $R_0 \rightarrow 0$ limit, we get that
\begin{equation}
\label{K0}
K_0(r,z) = r^2 k_d^{-\frac{2}{d-3}} F(r,z)^{\frac{2}{d-3}} \ .
\end{equation}
This result will be important below, since the
solution to the equations for $K(R,v)$ can be thought of
as a correction to $K_0(r,z)$ in \eqref{K0}. Thus, \eqref{K0}
is the zeroth order part of $K(r,z)$ and below we find
the leading correction to $K(r,z)$.

Notice that using \eqref{K0} we can find the flat space limit $R_0 \rightarrow
0$ of the black hole on cylinder solutions. 
Using the definition \eqref{defAK} we see that $\CM^d \times S^1$
has the flat space metric
\begin{equation}
ds^2 = - dt^2 + A_0 dR^2 + \frac{A_0}{K_0^{d-2}} dv^2 
+ K_0 R^2 d\Omega_{d-2}^2 \ .
\end{equation}
Comparing this with \eqref{cylcoord} we see that
\begin{equation}
\label{defR}
R^{d-3} = \frac{k_d}{F(r,z)}  \ .
\end{equation}
From requiring a diagonal metric in the $(R,v)$ coordinates
it is not hard to show that the resulting integrability condition
on $v$ is solved by \cite{Harmark:2002tr}
\begin{equation}
\label{defv}
\partial_r v = \frac{r^{d-2} }{(d-3) k_d} \partial_z F
\spa
\partial_z v = - \frac{r^{d-2} }{(d-3) k_d} \partial_r F \ .
\end{equation}
This in turn gives 
\begin{equation}
\label{A0}
A_0(r,z) = (d-3)^2 k_d^{-\frac{2}{d-3}}
\frac{F(r,z)^{2\frac{d-2}{d-3}}}{(\partial_r F)^2 + (\partial_z F)^2} \ .
\end{equation}
Note that both $A_0(R,v)$ and $K_0(R,v)$ are periodic in $v$ with
period $2\pi$.
We note that the above flat-space coordinate system 
is precisely that proposed in \cite{Harmark:2002tr}
for the flat-space limit of black holes on cylinders.

Finally, we note that using \eqref{K0}, \eqref{ansatz}, \eqref{gtt} 
and \eqref{nt1} we see that
\eqref{K0} in fact has the consequence that
\begin{equation}
\label{genFM}
\Phi(r,z) + B(r,z) = - \left( 1 - \frac{n}{d-2} \right) 
\frac{8 \pi G_{\rm N} M}{(d-1)\Omega_{d-1}} F(r,z) \ ,
\end{equation}
also for finite masses, which means that given a black
hole solution with a mass $M$ and binding energy $n$ we
can use \eqref{genFM} to find the Fourier modes $\varrho_k$.
This ensures the uniqueness of the black hole branch.%
\footnote{This observation is considered from another point of
view in \cite{Harmark:new}.}

\section{Ansatz in new coordinate system}
\label{secnewcoords}

In this section we define a new set of coordinates based
on the $(R,v)$ coordinates defined by the ansatz \eqref{ansatz}.
As we explain in the following, these new coordinates are
very useful to describe the metric near the horizon of a black
hole on a cylinder.

Consider a small black hole on a cylinder $\R^{d-1} \times S^1$.
We can think of this black hole as a one-dimensional array of black holes
in $\R^d$, the covering space for $\R^{d-1} \times S^1$. 
If we make the size of the black holes very small the
metric near a particular black hole in the array should be
like a $(d+1)$-dimensional Schwarzschild black hole.
The metric for a $(d+1)$-dimensional Schwarzschild black hole can
be written
\begin{equation}
\label{schw}
ds^2 = - \left( 1 - \frac{\rho_s^{d-2}}{\rho^{d-2}} \right) dt^2
+ \left( 1 - \frac{\rho_s^{d-2}}{\rho^{d-2}} \right)^{-1} d\rho^2
+ \rho^2 \left( d\theta^2 + \sin^2 \theta d\Omega_{d-2}^2 \right) \ .
\end{equation}
We have written out the $S^{d-1}$ part in an angle and an $S^{d-2}$
part since we have a general $SO(d-1)$ symmetry of our small black hole
solutions.
We now want to construct a new ansatz for small black holes
that asymptotes to the metric \eqref{schw} near the horizon
as $M \rightarrow 0$.

To do this, we first notice that the flat-space limit of 
the $(d+1)$-dimensional Schwarzschild black hole metric \eqref{schw}
is the spherical coordinate system on $\CM^{d+1}$ with metric
\begin{equation}
ds^2 = - dt^2 + d\rho^2 + 
\rho^2 d\theta^2 + \rho^2 \sin^2 \theta d\Omega_{d-2}^2 \ .
\end{equation}
We can also use this coordinate system for $\CM^d \times S^1$
if only we remember that $\CM^{d+1}$ is the covering space.
We can therefore relate the spherical coordinates $(\rho,\theta)$
to the cylindrical coordinates $(r,z)$ defined via the metric
\eqref{cylcoord} by the relations
\begin{equation}
\label{rhotheta}
r = \rho \sin \theta
\spa
z = \rho \cos \theta \ .
\end{equation}
Note here that the $(r,z)=(0,0)$ point, where the small black hole
singularity is located, corresponds to $\rho = 0$ in the spherical
coordinates.

We now want to define the new coordinates $\rht$ and $\tht$
in terms of the $(R,v)$ coordinates
so that $\rht=\rht(R)$ and $\tht=\tht(v)$ along
with the condition 
that $\rht/\rho \rightarrow 1$ and  $\tht/\theta \rightarrow 1$
for $R \rightarrow 0$ with $R_0 = 0$. 
It is not hard to see that all these requirements
are met by defining $(\rht,\tht)$ from
$(R,v)$ according to the relations
\begin{equation}
\label{defrtht}
R^{d-3} = k_d \rht^{d-2} \spa
v = \pi - \frac{d-2}{d-3} k_d^{-1} \int_{x=0}^{\tht} dx 
(\sin x)^{d-2} \ .
\end{equation}
Note here that $\tht = 0$ corresponds to $v=\pi$ and 
$\tht = \pi$ to $v=-\pi$.

If we in addition define the two functions
$\tilde{A}(\rht,\tht)$ and $\tilde{K}(\rht,\tht)$
by
\begin{equation}
A = \frac{(d-3)^2}{(d-2)^2} (k_d \rht)^{-\frac{2}{d-3}} \tilde{A}
\spa
K = \sin^2 \tht (k_d \rht)^{-\frac{2}{d-3}} \tilde{K} \ ,
\end{equation}
one can check that the ansatz \eqref{ansatz} now can
be written in the $(\rht,\tht)$ coordinates as
\begin{equation}
\label{newansatz}
ds^2 = - f dt^2 + \frac{\tilde{A}}{f} d\rht^2 
+ \frac{\tilde{A}}{\tilde{K}^{d-2}} \rht^2 d\tht^2
+ \tilde{K} \rht^2 \sin^2 \tht d\Omega_{d-2}^2
\spa
f = 1 - \frac{\rho_0^{d-2}}{\rht^{d-2}} \ ,
\end{equation}
where $\rho_0^{d-2} = k_d^{-1} R_0^{d-3}$.

We review the thermodynamics of the ansatz \eqref{newansatz}
in Appendix \ref{apptherm}.

\subsubsection*{Flat space limit of $(\rht,\tht)$ coordinates}

Take now the flat space limit $\rho_0 \rightarrow 0$ limit of the metric
\eqref{newansatz}. This gives the metric
\begin{equation}
\label{newflat}
ds^2 = - dt^2 + \tilde{A}_0 d\rht^2 
+ \frac{\tilde{A}_0}{\tilde{K}_0^{d-2}} \rht^2 d\tht^2
+ \tilde{K}_0 \rht^2 \sin^2 \tht d\Omega_{d-2}^2 \ ,
\end{equation}
with
\begin{equation}
\tilde{A}_0 (\rht,\tht) = \lim_{R_0 \rightarrow 0} \tilde{A} (\rht,\tht)
\spa
\tilde{K}_0 (\rht,\tht) = \lim_{R_0 \rightarrow 0} \tilde{K} (\rht,\tht) \ .
\end{equation}
Using \eqref{defR}-\eqref{defv} together with \eqref{defrtht} we see that
the flat space coordinates $(\rht,\tht)$
in terms of the $(\rho,\theta)$ coordinates 
are given by
\begin{equation}
\label{defrht}
\rht^{d-2} = \frac{1}{F(\rho,\theta)} \ ,
\end{equation}
\begin{equation}
\label{deftht}
(\sin \tht)^{d-2} \partial_\rho \tht =
\frac{\rho^{d-3} }{d-2} (\sin \theta)^{d-2} \partial_\theta F
\spa
(\sin \tht)^{d-2} \partial_\theta \tht =
- \frac{\rho^{d-1} }{d-2} (\sin \theta)^{d-2} \partial_\rho F \ .
\end{equation}
Here $F(\rho,\theta)$ is the function $F(r,z)$ defined
in \eqref{defF} written in $(\rho,\theta)$ coordinates 
(defined in \eqref{rhotheta}). 

We now want to study the flat space metric \eqref{newflat}
near the point $\rht = 0$, i.e. for $\rht \ll 1$, since that is where
the black hole is located. 
Clearly, from \eqref{defrht}, $\rht \ll 1$ is equivalent
to $\rho \ll 1$. Thus, as a first step, we need to understand $F(\rho,\theta)$
for $\rho \ll 1$.
Expanding $F(\rho,\theta)$ for $\rho \ll 1$ we get%
\footnote{Here $\zeta(s)$ is the Riemann Zeta function defined
as $\zeta(s) = \sum_{m=1}^\infty m^{-s}$.}
\begin{equation}
\label{Fexp}
F(\rho,\theta) = \frac{1}{\rho^{d-2}}
+ \frac{2\zeta(d-2)}{(2\pi)^{d-2}}
+ \frac{\zeta(d)}{(2\pi)^{d}} (d-2) \left[ d \cos^2 \theta - 1 \right] \rho^2
+ \mathcal{O} (\rho^4) \ .
\end{equation}
One can now use the expansion \eqref{Fexp} 
of $F(\rho,\theta)$ for $\rho \ll 1$ to find the relation between
$(\rho,\theta)$ and $(\rht,\tht)$ for $\rht \ll 1$.
We get%
\footnote{Note that \eqref{deftht} and \eqref{exptht} explicitly
shows that $\theta = 0$ is equivalent to $\tht =0$. In terms
of the $(R,v)$ coordinates this shows that $A_0(R,v)$
and $K_0(R,v)$ are periodic in $v$ with period $2\pi$, also near
the location of the black hole.}
\begin{equation}
\rho = \rht \left( 1 + \frac{2\zeta(d-2)}{(d-2) (2\pi)^{d-2}} \rht^{d-2}
+ \CO ( \rht^d ) \right) \ ,
\end{equation}
\begin{equation}
\label{exptht}
\sin^2 \theta = \sin^2 \tht \left( 1 + \frac{4 \zeta(d)}{(2\pi)^d}
\cos^2 \tht \rht^d + \CO (\rht^{d+2} ) \right) \ .
\end{equation}
One can easily obtain the higher order corrections as well.
However, those will not be of importance in this paper.

Finally, we are ready to find the expansions of $\tilde{A}_0(\rht,\tht)$
and $\tilde{K}_0(\rht,\tht)$ for $\rht \ll 1$.
From \eqref{K0} and \eqref{A0} along with \eqref{defrtht} we get
\begin{equation}
\tilde{K}_0 = \frac{\rho^2 \sin^2 \theta}{\rht^2 \sin^2 \tht}
\spa
\tilde{A}_0 = \left[ (\partial_\rho \rht)^2 
+ \rht^2 \tilde{K}_0^{-(d-2)} (\partial_\rho \tht)^2 \right]^{-1} \ .
\end{equation}
Using this with the expansions \eqref{exptht} we get
\begin{equation}
\label{tilA0}
\tilde{A}_0 (\rht,\tht) 
= 1 + \frac{4(d-1)\zeta(d-2)}{(d-2) (2\pi)^{d-2}} \rht^{d-2} 
+ \CO (\rht^d) \ ,
\end{equation}
\begin{equation}
\label{tilK0}
\tilde{K}_0 (\rht,\tht) 
= 1 + \frac{4\zeta(d-2)}{(d-2) (2\pi)^{d-2}} \rht^{d-2}
+ \CO (\rht^d) \ ,
\end{equation}
for $\rht \ll 1$. We included here the corrections to order $\rht^{d-2}$.
The following corrections at order $\rht^d$ depends on $\tht$.

\section{Corrected metric away from black hole}
\label{secNewt}

In this section we present the corrected metric for small
black holes for the region away from the black hole.
This is the region governed by the Newtonian limit of the
Einstein equations. This gives a first-order correction
to the flat-space metric that we use in Section \ref{secfull}
to construct the complete metric for small black holes.

In Section \ref{secn0sol} we find the general correction
to the metric for an arbitrary Newtonian gravity potential
in the ansatz \eqref{ansatz}.
In Section \ref{secinrt} we transform the result
of Section \ref{secn0sol} to the $(\rht,\tht)$ coordinates.

\subsection{Solving for general Newtonian gravity potential}
\label{secn0sol}

Consider the
Einstein equations for a general Newtonian gravity potential
$\Phi$ and a general binding energy potential $B$
\begin{eqnarray}
\label{einst}
& R^t_{\ t} = - \nabla^2 \Phi - \nabla^2 B \spa
R^z_{\ z} = \frac{1}{d-2} \nabla^2 \Phi + (d-2) \nabla^2 B \ ,
& \nn \\ &
R^r_{\ r} = R^{\phi_1}_{\ \phi_1} 
= \frac{1}{d-2} \nabla^2 \Phi - \nabla^2 B \ , &
\end{eqnarray}
where $\phi_1$ is one of the angles on the $S^{d-2}$ sphere.
The aim in the following is to find a solution of \eqref{einst}
for small masses, i.e. for $M\rightarrow 0$.
This can then subsequently be used to get the leading correction
in $R_0^{d-3}$ to the metric for small black holes.

In Section \ref{fourier} 
it is explained that $n \rightarrow 0$ for $M \rightarrow 0$.
This has the consequence that 
while $\Phi/(G_{\rm N}M)$ is finite for $M \rightarrow 0$
then $B/(G_{\rm N} M) \rightarrow 0$ for $M \rightarrow 0$. 
We see then from \eqref{einst} that we can neglect the 
$B$ potential since it is small compared to the $\Phi$ potential
for $M \rightarrow 0$.
In other words any correction to $n$ only appears as a second-order
effect in the correction of the metric. With respect to computing
the first order correction to the metric
we can therefore effectively set $n=0$.

Setting now $n=0$ we only have a Newtonian gravity potential
$\Phi$ and there are no potential for the binding energy.
In the $(R,v)$ coordinates this gives the Einstein equations
\begin{equation}
\label{einstn0}
R^t_{\ t} = - \nabla^2 \Phi
\spa
R^v_{\ v} = R^R_{\ R} = R^{\phi_1}_{\ \phi_1} = \frac{1}{d-2} \nabla^2 \Phi
\spa
R_{Rv} = 0 \ .
\end{equation}
We can moreover restrict ourselves to potentials $\Phi=\Phi(R)$ which
does not depend on $v$, since in the end we will put 
$\Phi=-\frac{1}{2}\frac{R_0^{d-3}}{R^{d-3}}$.
Note that then 
\begin{equation}
\nabla^2 \Phi = \frac{1}{A_0} \left( \Phi'' + \frac{d-2}{R} \Phi' \right) \ ,
\end{equation}
where prime refers to the derivative with respect to $R$.
We now want to solve the Einstein equations \eqref{einstn0}
to first order in $G_{\rm N}$.

The ansatz for the metric is
\begin{eqnarray}
\label{genans}
ds^2 &=& - \Big(1+2\Phi \Big) dt^2
+ \Big( 1 - 2 u + 2g \Big) A_{0} dR^2
+ \Big( 1 + 2g - (d-2) 2h \Big) \frac{A_{0}}{K_{0}^{d-2}} dv^2
\nn \\ && 
+ \Big( 1 + 2h \Big) K_{0} R^2 d\Omega_{d-2}^2 \ ,
\end{eqnarray}
where $u$, $g$ and $h$ are undetermined functions.
The ansatz \eqref{genans} is chosen so that for
$\Phi = u = - \frac{1}{2} \frac{R_0^{d-3}}{R^{d-3}}$ 
it reduces to a form consistent with the general ansatz
\eqref{ansatz}.
The idea is now to find $u$, $g$ and $h$ as functions of $\Phi$ and
$\Phi'$ so that the Einstein equations \eqref{einstn0}
are satisfied to first order.

Since $\partial_R^2 g$ is present in $R^R_{\ R}$ but 
$\partial_R^2 u$ and $\partial_R^2 h$ are not
we see that $g = g_1 \Phi$ since otherwise we will have a $\Phi'''$ 
term in $R^R_{\ R}$ which cannot be canceled by other terms. 
Similarly, since $R^v_{\ v}$ have $\partial_R^2 g$ and $\partial_R^2 h$ 
terms but not a $\partial_R^2 u$ term
we need that $h = h_1 \Phi$. Thus, our ansatz for $u$, $g$ and $h$ is
\begin{equation}
\label{ugh}
u = u_1 \Phi - (1-u_1) \frac{R}{d-3} \Phi'
\spa
g = g_1 \Phi 
\spa
h = h_1 \Phi \ .
\end{equation}
Note that we use the above ansatz for 
$u$ to ensure that $u = \Phi$ whenever 
$\Phi = -\frac{1}{2} \frac{R_0^{d-3}}{R^{d-3}}$.

After various algebraic manipulations we find
that the solution to the Einstein equations \eqref{einstn0} 
to first order is 
\begin{equation}
u = - \frac{R}{d-3} \Phi'
\spa
g = \frac{1}{d-3} \left( \frac{1}{d-2} 
+ \frac{R}{2} \frac{\partial_R A_0}{A_0}  \right) \Phi
\spa
h = \frac{1}{d-3} \left( \frac{1}{d-2} 
+ \frac{R}{2} \frac{\partial_R K_0} {K_0} \right) \Phi \ .
\end{equation}
Note that the $R \rightarrow 0$ and $R \rightarrow \infty$ limits
reproduces the results found previously in \cite{Harmark:2002tr}.

Putting then $\Phi = -\frac{1}{2} \frac{R_0^{d-3}}{R^{d-3}}$ we 
find that the leading correction in $R_0^{d-3}$ 
to the small black hole metric in the ansatz \eqref{ansatz}
is given by
\begin{equation}
\label{corA}
A = \left( 1 - \frac{1}{(d-2)(d-3)} \frac{R_0^{d-3}}{R^{d-3}}\right) A_0
- \frac{R}{2(d-3)} \frac{R_0^{d-3}}{R^{d-3}} \partial_R A_0 \ ,
\end{equation}
\begin{equation}
\label{corK}
K = \left( 1 - \frac{1}{(d-2)(d-3)} \frac{R_0^{d-3}}{R^{d-3}}\right) K_0
- \frac{R}{2(d-3)} \frac{R_0^{d-3}}{R^{d-3}} \partial_R K_0 \ .
\end{equation}
In conclusion, \eqref{corA}-\eqref{corK} describes the black
hole metric in the ansatz \eqref{ansatz} for $R \gg R_0$
when $R_0 \ll 1$.

\subsection{Corrected metric in $(\rht,\tht)$ coordinates}
\label{secinrt}

The leading correction \eqref{corA}-\eqref{corK} 
is easily transformed to the $(\rht,\tht)$ coordinates.
This gives
\begin{equation}
\label{AKfirst}
\tilde{A} = \tilde{A}_0 - 
\frac{\rht}{2(d-2)} \frac{\rho_0^{d-2}}{\rht^{d-2}}
\partial_{\rht} \tilde{A}_0
\spa
\tilde{K} = \tilde{K}_0 - 
\frac{\rht}{2(d-2)} \frac{\rho_0^{d-2}}{\rht^{d-2}}
\partial_{\rht} \tilde{K}_0 \ .
\end{equation}
Therefore, \eqref{AKfirst} describes the metric
(in the ansatz \eqref{newansatz}) for small black holes on cylinders
for $\rht \gg \rho_0$ to first order in $\rho_0^{d-2}$ when
$\rho_0 \ll 1$.

Using now the $\rht \ll 1$ expansion of $\tilde{A}_0$ and
$\tilde{K}_0$ found in \eqref{tilA0}-\eqref{tilK0} 
we get 
\begin{equation}
\label{cortA}
\tilde{A} = 1+ 
\frac{2(d-1)\zeta(d-2)}{(d-2) (2\pi)^{d-2}} 
\left[ 2\rht^{d-2} - \rho_0^{d-2} \right] + \CO ( \rht^d ) \ ,
\end{equation}
\begin{equation}
\label{cortK}
\tilde{K} = 1+ 
\frac{2\zeta(d-2)}{(d-2) (2\pi)^{d-2}} 
\left[ 2\rht^{d-2} - \rho_0^{d-2} \right] + \CO ( \rht^d ) \ .
\end{equation}
Thus,  the functions
\eqref{cortA}-\eqref{cortK} describes the metric
in the ansatz \eqref{newansatz} for $\rho_0 \ll \rht \ll 1$.
This result is used below to find the metric
of small black holes for $\rho_0 \leq \rht \ll 1$.

\section{Metric for small black holes on cylinders}
\label{secfull}

In this section we find the metric for small black holes
on cylinders.

In Section \ref{secNewt} we found that for $\rho_0 \ll \rht \ll 1$
the small black hole is described by \eqref{cortA}-\eqref{cortK}
in the ansatz \eqref{newansatz}. 

We now want to solve the vacuum Einstein equations for
$\rho_0 \leq \rho \ll 1$.
We first notice that the functions $\tilde{A}$ and $\tilde{K}$
given by \eqref{cortA}-\eqref{cortK} are independent of $\tht$.
This means that we can take $\tilde{A}$ and $\tilde{K}$ to be 
independent of $\tht$ for $\rho_0 \leq \rht \ll 1$. 
This can easily be argued using a systematic expansion
in terms of $\rho_0^{d-2}/\rht^{d-2}$.

Using now the result of Appendix \ref{appsph} that the metric 
\eqref{moregen} has the vacuum solutions given by 
the metric \eqref{umet} with the function $u$ given by \eqref{ures},
we get that for $\rho_0 \leq \rht \ll 1$ the functions $\tilde{A}$
and $\tilde{K}$ in the ansatz \eqref{newansatz} are given by
\begin{equation}
\label{AKw}
\tilde{A}^{- \frac{d-2}{2(d-1)}} 
= \tilde{K}^{- \frac{d-2}{2}} = 
\frac{1-w^2}{w} \frac{\rht^{d-2}}{\rho_0^{d-2}} + w \ ,
\end{equation}
with $w$ being a constant.
Comparing \eqref{AKw} with \eqref{cortA}-\eqref{cortK}
we see then that 
\begin{equation}
\label{thew}
w = 1 + \frac{\zeta(d-2)}{(2\pi)^{d-2}} \rho_0^{d-2}
+ \CO (\rho_0^{2(d-2)} ) \ .
\end{equation}
In conclusion, the metric of a small black hole on a cylinder
for $\rho_0 \leq \rht \ll 1$
is given by the ansatz \eqref{newansatz} with
the functions $\tilde{A}$ and $\tilde{K}$ given
as in \eqref{AKw} and \eqref{thew}.

We remind the reader that the metric for larger $\rht$ is given
by \eqref{AKfirst} in the ansatz \eqref{newansatz}. 
Thus, we have found the complete
metric, i.e. for all $\rht \geq \rho_0$, 
for black holes on cylinders with $\rho_0 \ll 1$
to order $\rho_0^{d-2}$. 
Since $M \propto \rho_0^{d-2}$ this means that we have
found the complete metric for small black holes on cylinders
to first order in the mass.

Summarizing the main result: For $\rho_0 \leq \rht \ll 1$
the metric of a small black hole on a cylinder $\R^{d-1} \times S^1$ is
given by
\begin{equation}
\label{met1}
ds^2 = - f dt^2 + f^{-1} G^{-\frac{2(d-1)}{d-2}} d\rht^2 
+ G^{-\frac{2}{d-2}} \rht^2 
\left( d\tht^2 + \sin^2 \tht \,  d\Omega_{d-2}^2 \right)
 \ ,
\end{equation}
\begin{equation}
\label{met2}
f = 1 - \frac{\rho_0^{d-2}}{\rht^{d-2}}
\spa
G(\rht) = \frac{1-w^2}{w} \frac{\rht^{d-2}}{\rho_0^{d-2}} + w 
\spa
w = 1 + \frac{\zeta(d-2)}{(2\pi)^{d-2}} \rho_0^{d-2}
+ \CO ( \rho_0^{2(d-2)} ) \ ,
\end{equation}
to first order in $\rho_0^{d-2}$.

We notice that $w=1$ in the metric \eqref{met1}-\eqref{met2}
corresponds to the $d+1$ dimensional Schwarzschild black hole metric
\eqref{schw}, 
thus we indeed
get that for $\rho_0 \rightarrow 0$ the small black hole
asymptotes to the $d+1$ dimensional Schwarzschild black hole.
Moreover, the $(\rht,\tht)$ coordinates asymptotes to the $(\rho,\theta)$
coordinates in this limit as expected.

In the rest of the paper we consider the consequences of the
small black hole on cylinder metric \eqref{met1}-\eqref{met2}
that we have obtained.

\section{Corrected thermodynamics}
\label{cortherm}

In this section we find the corrected thermodynamics that results
from the metric for small black holes on cylinder \eqref{met1}-\eqref{met2}.
Note that the general thermodynamics for the ansatz 
\eqref{newansatz} in $(\rht,\tht)$
coordinates is listed in Appendix \ref{apptherm}.

It is easily seen from \eqref{AKw} and \eqref{thew} that 
$\tilde{A}(\rht,\tht)$
on the horizon, as defined in \eqref{defAth}, is
\begin{equation}
\label{finalA}
\tilde{A}_h = 1 + \frac{2(d-1) \zeta(d-2)}{(d-2)(2\pi)^{d-2}} \rho_0^{d-2}
+ \CO ( \rho_0^{2(d-2)} ) \ .
\end{equation}
From this we can find the leading correction to the relative
binding energy. This is done using \eqref{relAn}. We get
\begin{equation}
\label{firstn}
n = \frac{(d-2)\zeta(d-2)}{2(2\pi)^{d-2}} \rho_0^{d-2}
+ \CO ( \rho_0^{2(d-2)} ) \ .
\end{equation}
We thus see that the relative binding energy $n$ is non-zero.
That $n$ becomes positive and not negative is expected since 
one cannot have negative $n$ \cite{Traschen:2003jm,Shiromizu:2003gc}
As we shall see below the fact that $n$ is non-zero signals
that the physics of black holes on cylinders is different from 
that of black holes in flat space.

Using \eqref{finalA} and \eqref{firstn} in \eqref{thermrho}
we now get the corrected thermodynamics
\begin{equation}
\label{corM}
M = \frac{(d-1)\Omega_{d-1}}{16\pi G_{\rm N}} \rho_0^{d-2} 
\left( 1 + \frac{\zeta(d-2)}{2(2\pi)^{d-2}} \rho_0^{d-2} 
+ \CO ( \rho_0^{2(d-2)} ) \right) \ ,
\end{equation}
\begin{equation}
T = \frac{d-2}{4\pi \rho_0} 
\left( 1 - \frac{(d-1)\zeta(d-2)}{(d-2)(2\pi)^{d-2}} \rho_0^{d-2} 
+ \CO ( \rho_0^{2(d-2)} ) \right) \ ,
\end{equation}
\begin{equation}
\label{corS}
S = \frac{\Omega_{d-1}}{4 G_{\rm N}} \rho_0^{d-1} 
\left( 1 + \frac{(d-1)\zeta(d-2)}{(d-2)(2\pi)^{d-2}} \rho_0^{d-2} 
+ \CO ( \rho_0^{2(d-2)} ) \right) \ .
\end{equation}
One can easily check that both the Smarr formula 
\cite{Harmark:2003dg,Kol:2003if} 
$(d-1)TS = (d-2-n)M$ 
and the first law of thermodynamics $\delta M = T \delta S$ holds.

We see that the corrected thermodynamics \eqref{corM}-\eqref{corS}
becomes increasingly like
that of a Schwarzchild black hole in a $d+1$ space-time as 
$M \rightarrow 0$, exactly as one would expect. The corrections
\eqref{corM}-\eqref{corS} thus encapture the departure
of the thermodynamics
of black holes on cylinders from that of 
the Schwarzchild black hole.

To see more clearly
what this means for the thermodynamics, we can compute
\begin{equation}
\label{logSlogM1}
\frac{\delta \log S}{\delta \log M} = \frac{d-1}{d-2} 
\left( 1 + \frac{\zeta(d-2)}{2(2\pi)^{d-2}} \rho_0^{d-2} 
+ \CO ( \rho_0^{2(d-2)} ) \right) \ .
\end{equation}
This shows explicitly that the thermodynamic nature of the 
black hole changes as we start increasing the mass, since
$\delta \log S/ \delta \log M$ clearly characterizes the thermodynamics.
That $\delta \log S/ \delta \log M$ increases can be understod from the
general formula \cite{Harmark:2003dg}
\begin{equation}
\label{genSM}
\frac{\delta \log S}{\delta \log M} = \frac{d-1}{d-2-n} \ ,
\end{equation}
and the fact that $n > 0$.

As advocated in \cite{Harmark:2003dg,Harmark:2003eg} 
it is useful to depict the black hole
branch in an $(M,n)$ diagram. It is therefore interesting to
find $n$ as function of $M$. However, instead of $M$ it is useful
to use
the dimensionless parameter
\begin{equation}
\label{defmu}
\mu = \frac{16 \pi G_{\rm N} M}{L^{d-2}} 
= \frac{16 \pi G_{\rm N} M}{(2\pi)^{d-2}} \ .
\end{equation}
Here we have used that the circumference of the circle
is $L=2\pi$ in our units.
We then get 
\begin{equation}
\label{nofM}
n = \frac{(d-2)\zeta(d-2)}{2(d-1)\Omega_{d-1}} \, \mu + \CO ( \mu^2 ) \ .
\end{equation}
Using \eqref{genSM} we get furthermore
\begin{equation}
\label{logSlogM2}
\frac{\delta \log S}{\delta \log \mu} = \frac{d-1}{d-2} 
\left( 1 +  \frac{\zeta(d-2)}{2(d-1)\Omega_{d-1}} \, \mu
+ \CO ( \mu^2 ) \right) \ .
\end{equation}
The result \eqref{genSM} in a sense encaptures
all of the corrected thermodynamics for black holes on cylinders
since we can integrate this relation.
For completeness we write here that the result of the integration
is
\begin{equation}
S ( \mu ) = c_1 \mu^{\frac{d-1}{d-2}} 
\left( 1 +  \frac{\zeta(d-2)}{2(d-2)\Omega_{d-1}} \,  \mu
 + \CO ( \mu^2 )\right)
\end{equation}
where $c_1$ is a constant given in \eqref{thecs}.

\subsubsection*{Thermodynamics for black hole copies}

In \cite{Harmark:2003eg} copies of the black hole branch
were introduced, based on an idea of Horowitz \cite{Horowitz:2002dc}.
The $k$'th black hole copy of the black hole is the solution
one gets by putting $k$ black holes along the circle 
direction of the cylinder, with all $k$ black holes
having an equal distance to each other.
Physically it is clear that all these copies are
unstable and have less entropy the higher $k$ gets.
Indeed, this was found to be the case for the leading
order thermodynamics of black holes and the copies
in \cite{Harmark:2003eg}. However, it is not a priori
clear that that continues to be valid for the corrected 
thermodynamics \eqref{corM}-\eqref{corS}.

Denoting the entropy of the $k$'th copy of the 
black hole branch as $S_k$ we get using \eqref{logSlogM2}
that
\begin{equation}
\log S_k(\mu) = \log c_k + \frac{d-1}{d-2} \log \mu + a_k \mu
\ ,
\end{equation}
\begin{equation}
\label{thecs}
\log c_k = \log c_1 - \frac{1}{d-2} \log k 
\spa
c_1 = \frac{(2\pi)^{d-1}}{4 (d-1)^{\frac{d-1}{d-2}} \Omega_{d-1}^{\frac{1}{d-2}}} \ ,
\end{equation}
\begin{equation}
a_k = k^{d-3} a_1 \spa
a_1 = \frac{\zeta(d-2)}{2(d-2)\Omega_{d-1}} \ .
\end{equation}
Physically we expect then that 
$S_k (\mu ) < S_{k'} (\mu)$ if $k > k'$.

If we now consider two copies $k' < k$ we find
\begin{equation}
\log S_k (\mu) > \log S_{k'} (\mu)
\Leftrightarrow 
\mu > \frac{1}{(d-2)a_1} 
\frac{\log k' - \log k}{k^{d-3}-(k')^{d-3}}  \ .
\end{equation}
Now, if we call $\mu_{\rm max}$ the maximally allowed $\mu$ 
 for the formula \eqref{logSlogM2}
to be approximately correct, we can see how large $\mu_{\rm max}$
has to be in order for the counter-intuitive situation
$\log S_k (\mu) > \log S_{k'} (\mu)$ to occur. This happens
if
\begin{equation}
\mu_{\rm max} > \frac{1}{(d-1)a_1} 
\frac{-\log (k'/k)}{1-\left( \frac{k'}{k} \right)^{d-3}} \ .
\end{equation}
The minimum of the right-hand side is at $k'/k \rightarrow 1$, 
so we get that
\begin{equation}
\mu_{\rm max} > \frac{2\Omega_{d-1}}{(d-3)\zeta(d-2)}  \ .
\end{equation}
For $d=5$ this gives $\mu_{\rm max} > 22$ which seems unreasonably
large as one also can see from the considerations in Section
\ref{secdiag}.
The corrected thermodynamics \eqref{logSlogM2} is thus not 
in contradiction with the expected physical properties
of the black holes copies.

\section{Phase diagram for black holes and black strings}
\label{secdiag}

As mentioned in the introduction one of the motivations
to find the metric for black holes on cylinders is to
get a better understanding of the phase structure of 
black objects, e.g. black holes and strings, on the cylinder.
As advocated in \cite{Harmark:2003dg,Harmark:2003eg} 
it is useful to draw the $(M,n)$ phase diagram
in order to understand this phase structure. 

\begin{figure}[ht]
\centerline{\epsfig{file=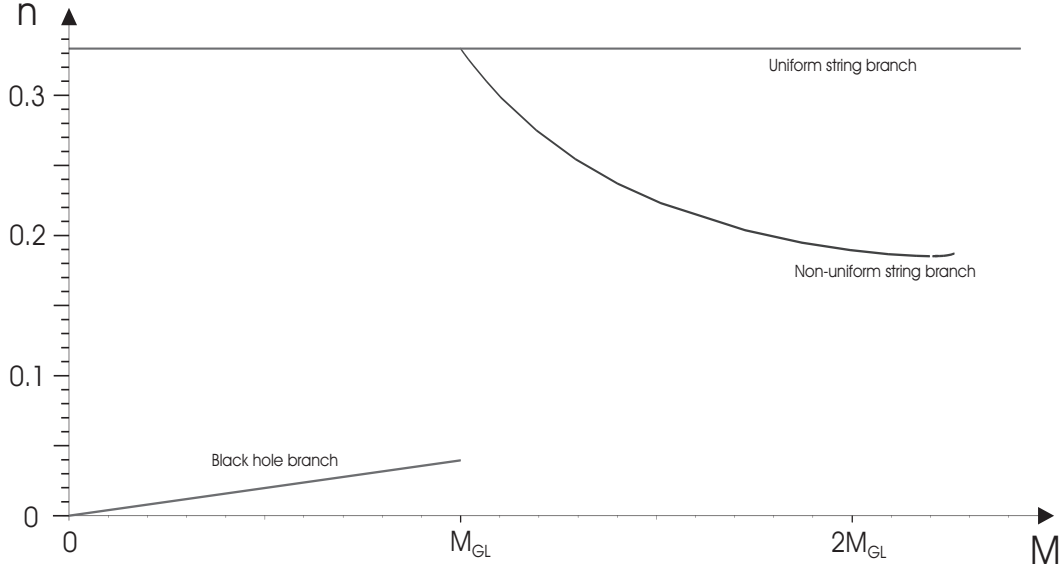,width=14cm,height=7.5cm}}
\caption{$(M,n)$ phase diagram for $d=5$ containing
the black hole branch, the
uniform string branch and the non-uniform string branch
of Wiseman.}
\label{Mndiag}
\end{figure}

In \cite{Harmark:2003dg,Harmark:2003eg} 
the $(M,n)$ phase diagram for the $d=5$ case was drawn.
The phase diagram included the uniform black string branch and
the non-uniform black string branch found numerically 
by Wiseman \cite{Wiseman:2002zc}.
Using our results on small black holes, we can now
put in part of the black hole branch in this $(M,n)$ diagram.
From \eqref{nofM} we compute that for $d=5$
\begin{equation}
\label{nplot}
n \simeq 0.017 \, \mu \simeq 0.040 \, \frac{M}{M_{\rm GL}} \ .
\end{equation}
Here we also listed $n$ as function of $M/M_{\rm GL}$ using
that the Gregory-Laflamme mass is $\mu_{\rm GL} = 2.31$ for $d=5$
\cite{Gregory:1993vy,Gregory:1994bj} (see \cite{Harmark:2003dg}
for the explicit numerical values of $\mu_{\rm GL}$).
Using \eqref{nplot} we have depicted the three known
phases of black objects in Figure \ref{Mndiag}.

A comment is in order here.
In Figure \ref{Mndiag} we have continued the linear behavior of 
$n$ as function of $M$ all the way up to $M= M_{\rm GL}$.
To see why we expect \eqref{nplot} to be approximately valid
up to $M= \, M_{\rm GL}$, we consider the function $F(\rho,\theta)$
in \eqref{Fexp}. We have included the first correction in 
$F(\rho,\theta)$ but not the second one. 
The value of $\rho$ for which the second correction is of equal
size as the first correction is given by
$\rho^2 = 8 \pi^2 \zeta(d-2)/ ((d-1)(d-2)\zeta(d))$.
For $d=5$ this gives that not including the
second term in $F(\rho,\theta)$ is a good approximation for
$\rho \ll 2.8$. In terms of the horizon radius this means
that $\rho_0 \ll 2.8$.
Thus, we see that \eqref{nplot} should be valid
for $\rho_0^3 \ll 21$ which translates to $\mu \ll 9$
and furthermore to $M \ll 4 \, M_{\rm GL}$.
This therefore makes it likely that \eqref{nplot}
is valid to a good approximation up to $M \simeq M_{\rm GL}$.

\section{Discussion and conclusions}
\label{secconcl}

The main results of this paper are:
\begin{itemize}
\item We have found the complete metric for small
black holes on cylinders $\R^{d-1} \times S^1$. 
For $\tilde{\rho} \gg \rho_0$ the metric is given
by \eqref{AKfirst} while for $\rho_0 \leq \tilde{\rho} \ll 1$
the metric is given by \eqref{met1}-\eqref{met2}.
The metric is valid to first order in $\rho_0^{d-2}$,
which means to first order in the mass.
\item We have found the corrected thermodynamics
using the metric \eqref{met1}-\eqref{met2}.
The corrected thermodynamics is \eqref{corM}-\eqref{corS}.
We can summarize the corrected thermodynamics in the formula
\eqref{logSlogM2}
\begin{equation*}
\frac{\delta \log S}{\delta \log \mu} = \frac{d-1}{d-2} 
\left( 1 +  \frac{\zeta(d-2)}{2(d-1)\Omega_{d-1}} \, \mu
+ \CO ( \mu^2 ) \right) \ ,
\end{equation*}
where $\mu$ is the rescaled mass in \eqref{defmu}.
\item We obtained in \eqref{nofM}
the corrected relative binding energy $n$ which
we found to increase when increasing the mass. 
If we allow variations of the circumference $L$
of the cylinder, the first law of thermodynamics is
$\delta M = T \delta S + nM L^{-1} \delta L$ \cite{Harmark:2003eg,Kol:2003if}.
Therefore, a non-zero $n$ means that the black hole
does not behave point-like (a point-like object
would have $\delta M = 0$ under a variation of $L$).
Qualitatively, this means that the physics of black
holes on cylinders are governed by the shape of the event horizon
rather than that of the singularity.
\end{itemize}

The fact that we were able to find the complete
metric describing small black holes can be taken
as a confirmation on a basic assumption about
the nature of black holes:
That black holes obey the principle of locality.

The ``principle of locality for black holes''  
means here that we believe that sufficiently
small black holes should not be influenced by
the global structure of the space-time.
Thus, for any locally flat space-time 
it should be so that
small black holes behave like in flat space.
For a black hole on the cylinder $\R^{d-1} \times S^1$
this means that as the mass $M \rightarrow 0$
it become more and more 
like a $(d+1)$-dimensional Schwarzschild black hole.
Using for example \eqref{genSM} this assumption on black
holes on cylinders is seen to be equivalent to the assumption
that $n \rightarrow 0$ for $M \rightarrow 0$.
That we in this paper are able to find 
a metric for small black holes is thus a non-trivial
consistency check on this assumption. 

A connected assumption is the one of Section \ref{fourier} that
as the black hole become smaller it behaves more and more like
a point-like object. This assumption leads to the prediction \eqref{genFM}
of the Fourier modes for the complete black hole branch,
as seen in Section \ref{fourier}.
The argument of Section \ref{fourier} 
used that $n \rightarrow 0$ for $M \rightarrow 0$
and that Newtonian physics should take over most of the space-time
as $M \rightarrow 0$.
Again, that we found the complete metric for small black holes
is a non-trivial check on these arguments.
It seems therefore reasonable to expect that the complete
black hole branch have the Fourier modes given by 
\eqref{genFM}.

We have also seen that the ansatz \eqref{ansatz}
proposed in \cite{Harmark:2002tr} and proven in 
\cite{Wiseman:2002ti,Harmark:2003eg} 
is highly succesful in describing small black holes.
This paper can therefore be seen as a confirmation
on the usefulness of this ansatz.

The success of the methods of this paper makes it natural
to ask whether they can be continued and one can find
higher order corrections to black holes on cylinders.
We believe that indeed is the case. 

Obviously, finding more corrections would be highly
interesting in view of the ongoing discussion on which
scenario of the black hole/black string transitions
that is the correct one. That is unfortunately
still unclear, even after the numerical work on 
black holes on the cylinder
$\R^4 \times S^1$ in \cite{Sorkin:2003ka,Kudoh:2003ki}. 

Finally, we comment that the non-zero relative binding energy
that we found for small black holes on cylinders means that 
the so-called ``Uniqueness Hypothesis''
of \cite{Harmark:2003dg,Harmark:2003eg}, stating that there only exists
one neutral and static black object for a given $M$ and $n$, 
still seem to hold for black holes and strings on cylinders.
If the relative binding energy would have been zero the 
Uniqueness Hypothesis would be violated due to the black
hole copies (see \cite{Harmark:2003eg} and Section \ref{cortherm}).

\section*{Acknowledgments}

We thank Niels Obers for many useful discussions.


\begin{appendix}

\section{Thermodynamics in the $(R,v)$ and $(\rht,\tht)$
coordinates}
\label{apptherm}

\subsubsection*{Thermodynamics in the $(R,v)$ coordinates}

We review here the thermodynamics in terms
of the $(R,v)$ coordinates defined by the ansatz \eqref{ansatz}, as found in
\cite{Harmark:2002tr}.
Define
\begin{equation}
\label{defAh}
A_h \equiv A(R,v) |_{R=R_0} \ ,
\end{equation}
which, as shown in \cite{Harmark:2002tr}, is independent of $v$.
Let furthermore the asymptotic behavior of $K(R,v)$
for $R \rightarrow \infty$ be written as%
\footnote{Note here that with \eqref{defchi} as the behavior of
$K(R,v)$ for $R \rightarrow \infty$ we get that
$A(R,v) = 1 - \chi \frac{R_0^{d-3}}{R^{d-3}} + \CO ( R^{-2(d-3)} )$
from the equations of motion \cite{Harmark:2002tr}.}
\begin{equation}
\label{defchi}
K(R,v) = 1 - \chi \frac{R_0^{d-3}}{R^{d-3}} + \CO ( R^{-2(d-3)} ) \ ,
\end{equation}
then we have using \eqref{gtt}-\eqref{gzz} with \eqref{phirz}-\eqref{defh} 
\cite{Harmark:2002tr,Harmark:2003dg}
\begin{equation}
\label{nt1}
M = \frac{\Omega_{d-2}}{8 G_{\rm N}} R_0^{d-3}
\frac{(d-1)(d-3)}{d-2-n}
\spa
n = \frac{1-(d-2)(d-3)\chi}{d-2 - (d-3)\chi} \ ,
\end{equation}
\begin{equation}
\label{nt2}
T = \frac{d-3}{4\pi \sqrt{A_h}R_0} \spa
S = \frac{2\pi \Omega_{d-2}}{4 G_{\rm N}} \sqrt{A_h} R_0^{d-2} \ .
\end{equation}
\eqref{nt1}-\eqref{nt2} gives the thermodynamics
of solutions described in the ansatz \eqref{ansatz}.

It was derived in \cite{Harmark:2003eg,Kol:2003if}
that the first law of thermodynamics
\begin{equation}
\label{first}
\delta M = T \delta S \ ,
\end{equation}
holds for the black hole solutions. 
Obviously on the black hole branch we can consider all
the quantities as being functions of $R_0$ only.
Therefore, we can express \eqref{first} as 
$\delta M/\delta R_0 = T \delta S/\delta R_0$. 
Thus, in terms of the above
defined quantities $A_h$ and $n$ the first law is equivalent to
\begin{equation}
\label{dAdR0}
\frac{1}{2} \frac{R_0}{A_h} \frac{\delta A_h}{\delta R_0} 
= \frac{d-1}{(d-2-n)^2} R_0 \frac{\delta n}{\delta R_0}
- \frac{1 - (d-2)n}{d-2-n} \ .
\end{equation}
%

\subsubsection*{Thermodynamics in the $(\rht,\tht)$ coordinates}

We review here the thermodynamics thermodynamics in terms of
the $(\rht,\tht)$ coordinates defined by \eqref{newansatz}. 
Define
\begin{equation}
\label{defAth}
\tilde{A}_h \equiv \tilde{A}(\rht,\tht) |_{\rht=\rho_0} \ .
\end{equation}
The thermodynamics is then
\begin{equation}
\label{thermrho}
M = \frac{\Omega_{d-1}}{16\pi G_{\rm N}} \rho_0^{d-2} 
\frac{(d-1)(d-2)}{d-2-n}
\spa
T = \frac{d-2}{4\pi \rho_0 \sqrt{\tilde{A}_h}}
\spa
S = \frac{\Omega_{d-1}}{4 G_{\rm N}} \rho_0^{d-1} \sqrt{\tilde{A}_h} \ .
\end{equation}
We note that for $\tilde{A}_h = 1$ and $n=0$ 
this thermodynamics is that of a Schwarzschild black hole
in $d+1$ dimensions. 

The first law in the form \eqref{dAdR0} gives
\begin{equation}
\label{relAn}
\frac{1}{2} \rho_0 (\log \tilde{A}_h )'
= \frac{(d-1)n}{d-2-n} + \frac{d-1}{(d-2-n)^2} \rho_0 n' \ ,
\end{equation}
where the prime denotes the derivative with respect to $\rho_0$.
This relation is of importance in the text.

\section{Derivation of general spherical metric}
\label{appsph}

We start with a $d+1$ dimensional metric of the form
\begin{equation}
\label{moregen}
ds^2 = - f dt^2 + f^{-1} e^{q} d\rht^2 
+ e^{q-(d-2)u} \rht^2 d\tht^2 + e^u \rht^2 \sin^2 \tht d\Omega_{d-2}^2
\spa
f = 1 - \frac{\rho_0^{d-2}}{\rht^{d-2}} \ ,
\end{equation}
with $q=q(\rht)$ and $u=u(\rht)$, i.e.
without any $\tht$ dependence. 
We see that \eqref{moregen} is in the form of the ansatz
\eqref{newansatz} with $\tilde{A} = e^q$ and $\tilde{K}=e^u$.
We want to analyze the solutions of the vacuum Einstein equations
with the metric \eqref{moregen}.

From $R_{\rht \tht} = 0$ we immediately get that $q' = (d-1) u'$.
We can therefore write $q = s + (d-1)u$ where $s$ is a
constant.
Write now $d\Omega_{d-2}^2 = d\phi_1^2 + \sin^2 \phi_1 d\phi_2^2
+ \cdots + \sin^2 \phi_1 \cdots \sin^2 \phi_{d-3} d\phi_{d-2}^2$,
where $\phi_1,...,\phi_{d-2}$ are the angles of $S^{d-2}$.
If we consider the Einstein equation $R^{\phi_1}_{\ \phi_1}=0$
it is easy to see that that equation only can be satisfied provided
$s=0$.
Therefore, the metric \eqref{moregen} reduces to
\begin{equation}
\label{umet}
ds^2 = - f dt^2 + f^{-1} e^{(d-1)u} d\rht^2 
+ e^{u} \rht^2 \left( d\tht^2 + \sin^2 \tht d\Omega_{d-2}^2 \right)
\spa
f = 1 - \frac{\rho_0^{d-2}}{\rht^{d-2}} \ .
\end{equation}
We now consider the solutions of the vacuum Einstein equations
for this metric, still with $u=u(\rht)$.
The remaining non-trivial Einstein equations gives
the two equations
\begin{equation}
\label{ueq1}
u'' - \frac{d-3}{\rht} u' - \frac{d-2}{2} (u')^2 =0 \ ,
\end{equation}
\begin{equation}
\label{ueq2}
\left( 1 - \frac{\rho_0^{d-2}}{\rht^{d-2}} \right) u''
+ \frac{d-1}{\rht} u' - \frac{\rho_0^{d-2}}{\rht^{d-2}} \frac{1}{\rht} u'
- \frac{2 (d-2)}{\rht^2} \left( e^{(d-2)u} - 1 \right) = 0 \ .
\end{equation}
Defining 
\begin{equation}
G(\rht) = \exp \left( - \frac{d-2}{2} u(\rht) \right) \ ,
\end{equation}
we see that \eqref{ueq1} becomes
\begin{equation}
G'' - \frac{d-3}{\rht} G' = 0 \ .
\end{equation}
The most general solution to this equation is
\begin{equation}
\label{Ggen}
G = c_1 \rho^{d-2} + c_2 \ ,
\end{equation}
with $c_1$ and $c_2$ being arbitrary constants.
Putting \eqref{Ggen} into \eqref{ueq2} we get that
\eqref{ueq2} is fulfilled if and only if
\begin{equation}
\rho_0^{d-2} c_1 c_2 + c_2^2 = 1 \ .
\end{equation}
Thus, we get that the most general vacuum solution with
metric \eqref{umet} is given by
\begin{equation}
\label{ures}
\exp \left( - \frac{d-2}{2} u \right) = 
\frac{1-w^2}{w} \frac{\rht^{d-2}}{\rho_0^{d-2}} + w  \ ,
\end{equation}
where $w$ is an arbitrary constant.
This means that the most general vacuum solution with
metric \eqref{moregen} is given by
$q = (d-1) u$ and \eqref{ures}.

\end{appendix}

\addcontentsline{toc}{section}{References}


\providecommand{\href}[2]{#2}\begingroup\raggedright\endgroup

\end{document}